# Life without dUTPase

Csaba Kerepesi[3], Judit E. Szabó[2], Vince Grolmusz[3,4], Beáta G. Vértessy[1,2,*]

[1]Department of Applied Biotechnology and Food Sciences, Budapest University of Technology and Economics, Budapest, 1111, Hungary, [2]Institute of Enzymology, Research Centre for Natural Sciences, Hungarian Academy of Sciences, Budapest, 1117, Hungary and [3] PIT Bioinformatics Group, Mathematical Institute, Eötvös University, Budapest,1117, Hungary, [4] Uratim Ltd., Budapest, 1118, Hungary

* To whom correspondence should be addressed. Tel:+36 1 382 6707; Fax: +36 1 463 3855; Email: vertessy@mail.bme.hu, vertessy.beata@ttk.mta.hu

**Abstract**

Fine-tuned regulation of the cellular nucleotide pools is indispensable for faithful replication of DNA. The genetic information is also safeguarded by DNA damage recognition and repair processes. Uracil is one of the most frequently occurring erroneous base in DNA; it can arise from cytosine deamination or thymine-replacing incorporation. Two enzyme families are primarily involved in keeping DNA uracil-free: dUTPases that prevent thymine-replacing incorporation and uracil-DNA glycosylases that excise uracil from DNA and initiate uracil-excision repair. Both dUTPase and the most efficient uracil-DNA glycosylase UNG is thought to be ubiquitous in free-living organisms. In the present work, we have systematically investigated the genotype of deposited fully sequenced bacterial and Archaeal genomes. Surprisingly, we have found that in contrast to the generally held opinion, a wide number of bacterial and Archaeal species lack the dUTPase gene(s). The *dut-* genotype is present in diverse bacterial phyla indicating that loss of this (or these) gene(s) has occurred multiple times during evolution. We have identified several survival strategies in lack of dUTPases: i) simultaneous lack or inhibition of UNG, ii) acquisition of a less dUTP-specific sanitizing nucleotide pyrophosphatase, and iii) supply of dUTPase from bacteriophages. Our data indicate that several unicellular microorganisms may efficiently cope with a *dut-* genotype potentially leading to an unusual uracil-enrichment in their genomic DNA.

**Funding**

Hungarian Scientific Research Fund OTKA [NK 84008, K109486]; Baross Program of the New Hungary Development Plan [3DSTRUCT, OMFB-00266/2010 REG-KM- 09-1-2009-0050]; Hungarian Academy of Sciences ([TTK IF-28/ 2012]; MedinProt program); an ICGEB Research Grant to BGV and the European Commission FP7 Biostruct-X project [contract No. 283570].

## Introduction

The DNA macromolecule is the repository for genomic information in most organisms (with the notable exception of RNA viruses). Stable storage and faithful transmission of genomic information would optimally require a stable macromolecule for these roles. However, the inherent chemical reactivity of DNA and the presence of reactive metabolites and other molecular species within the cell leads to numerous chemical modifications within the DNA even under normal, physiological conditions (1-4). Mutations arising from these modifications need to be kept under control, and numerous DNA damage recognition and repair processes evolved to deal with these problems (5). It is also important to mention that mutations are important instruments in driving evolutionary changes and development, as well. Especially for single cell organisms, eminently for bacteria, increased mutational rates leading to new phenotypes may be even advantageous for the species – appearance of antibiotic resistant strains may be a prominent example in this respect (6,7). Meanwhile, cells that acquired mutations deleterious for the phenotype will be overgrown by cells with advantageous mutations. In multicellular eukaryotes, such evolutionary changes are more complex since, in these organisms, the viable phenotype is more restricted due to the highly increased interactions within the cellular environment and also with the other cells/organs.

In response to the need of conserving the DNA-encoded information, a number of specific and highly efficient DNA repair pathways have evolved, such as base-excision repair, nucleotide excision repair, mismatch repair and double-strand break repair (8). These are strongly conserved from bacteria to man, and the protein factors responsible for these processes are usually ubiquitous, although the cognate protein families and isoforms may differ among organisms of different evolutionary branches. For pathways of key significance, it is also frequently observed that multiple protein families with similar functions are present in one organism to safeguard DNA-encoded information (9). In addition to the dedicated DNA damage recognition and repair pathways, sanitization and proper balance of the nucleotide pools are also of high importance (10). Hence, regulation of nucleotide *de novo* biosynthesis and salvage pathways need to be fine-tuned, and unwanted dNTPs, such as dUTP and dITP have to be removed. Sanitizing enzymes are usually dNTPases catalyzing pyrophosphorolysis of the specific un-orthodox dNTPs (11). A prominent example in this regard is the dUTPase enzyme family, representatives of which are considered to be ubiquitous and essential for viability in all free-living organisms (3,12,13). There is an intimate cross-talk between enzymes responsible for sanitizing of nucleotide pools and the respective base-excision repair DNA N-glycosylases that act hand in hand first to prevent incorporation of the unwanted nucleotide building block containing modified bases into newly synthesizing DNA and second, to excise those moieties that escaped the preventive measure or got produced within the DNA *in situ*. For the uracil moiety, the preventive/excising enzyme activities are presented by the dUTPase and the uracil-DNA glycosylase enzyme families, respectively (12-16).

The crosstalk between preventive and excising activities constitutes joint functional efforts with the aim to guard genome integrity. For the dUTPase/UNG enzyme pair, knock-out of the preventive activity of dUTPase is highly dangerous for the cell because it induces numerous uracil-incorporation events that will overload the base excision repair mechanism and transforms it into a hyperactive futile cycle (12,13,17,18). Knock-out of UNG, however, can be tolerated (19). In an *ung-/-* background, complementing enzyme families with uracil-DNA excising activities (TDG/MUG, SMUG, MPD4 enzyme families) are still functional, although less effective (9,20). Also, organisms with uracil-substituted DNA are still viable in lack of UNG, the most efficient uracil-excising enzyme (13,21).

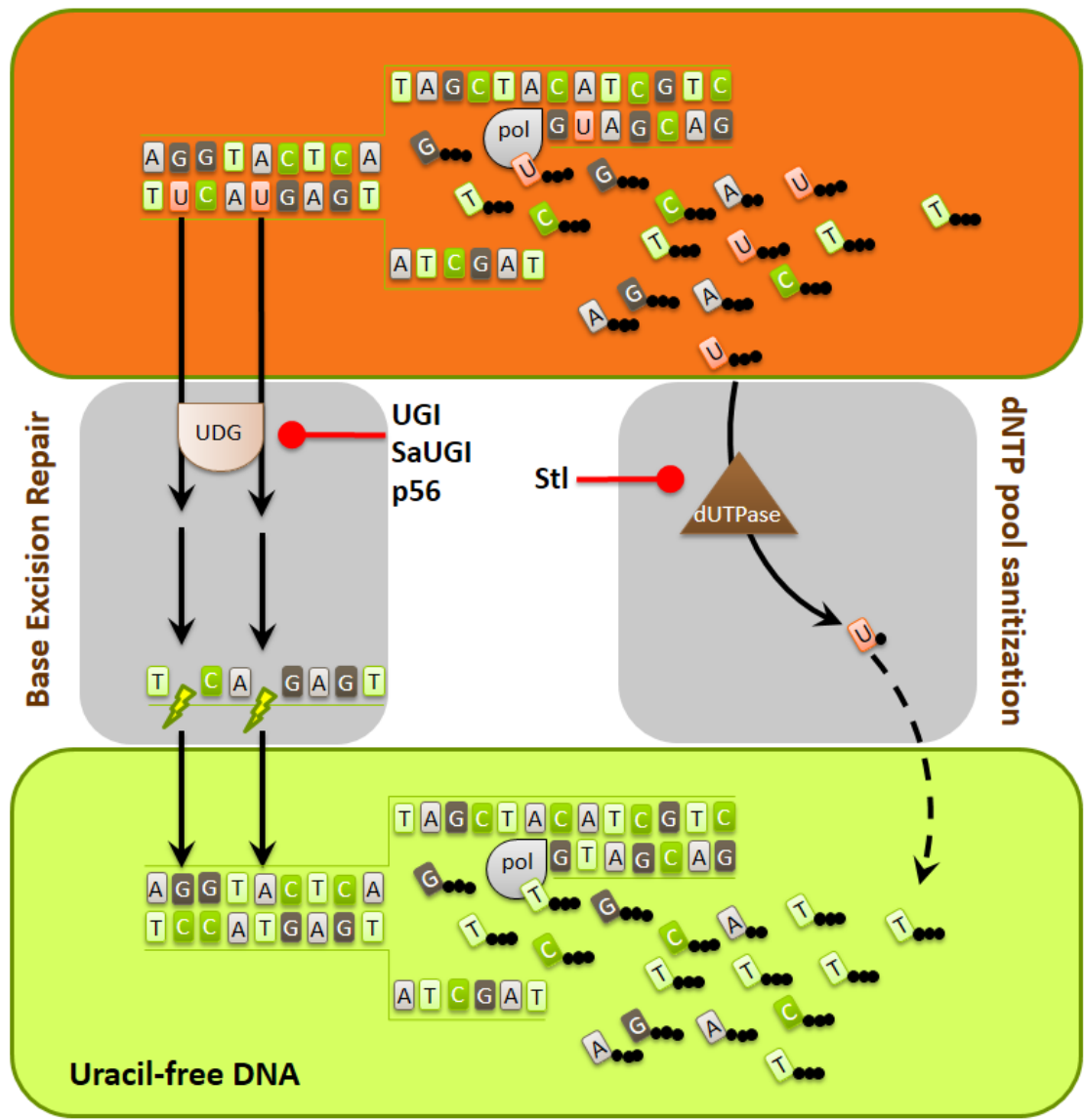

*Figure 1. Pathways and protein factors involved in the metabolism of uracil-substituted DNA. The scheme illustrates that dUTPase and UDG are responsible for keeping uracil out of DNA by dNTP pool sanitization or uracil-excision, respectively. Inhibitor proteins against UDG (UGI, SaUGI and p56) and dUTPase (Stl) are also included on the figure, showing their point of inhibitory attack.*

In a dUTPase knock-out background, viability can be still restored in some cases by simultaneous UNG knock-out (14,15,22), or by inhibiting the UNG enzyme with its specific and highly efficient protein inhibitor, UGI. In the double mutant organisms, the uracil content within DNA is highly elevated, however, the cells can survive, most probably since the majority of uracil moieties under these conditions are present as thymine-replacements, i.e., with the same Watson-Crick coding characteristics. Such circumstances have been observed in artificially engineered bacteria (*E. coli*), or similar situations are also found in specific life stages of wild type *Drosophila melanogaster* where dUTPase is down-regulated during development and the *ung* gene is absent from the genome (13,21).

However, to our knowledge, there is no report published on any free-living organism where the gene for dUTPase is not present within the genome. Our recent observations in several Staphylococcus strains shed light on circumstances where the dUTPase gene on the bacterial chromosome is present only due to insertion of a phage-encoded gene (in prophage form) (16). A wide survey of Staphylococcal strains also revealed several occasions where strains are viable and infectious in the absence of dUTPase gene(s) present in the genome, still, these strains are viable (23,24). This intriguing situation prompted us to investigate in details the genotypes of prokaryotes and Archaea with respect to the existence of genes primarily involved in uracil-DNA metabolism. Towards this aim, we have analyzed all fully-sequenced bacterial and Archaeal genomes deposited in NCBI, that is, 2261 bacterial and 151 Archaeal genomic sequence sets. In these investigations, we have specifically looked for the existence or lack of the genes of the dUTPase enzyme families, UNG the most proficient uracil-DNA glycosylase, as well as the genes for the proteins, described up to date as inhibitors of either dUTPase or UNG. Results clearly showed that numerous investigated microbes do not possess dUTPase genes, and this genotype can be paired with different patterns of presence/absence of UNG and inhibitor proteins. We conclude that the genetic distribution of proteins involved in uracil-DNA metabolism is unexpectedly diverse, and these conditions may have physiological consequences.

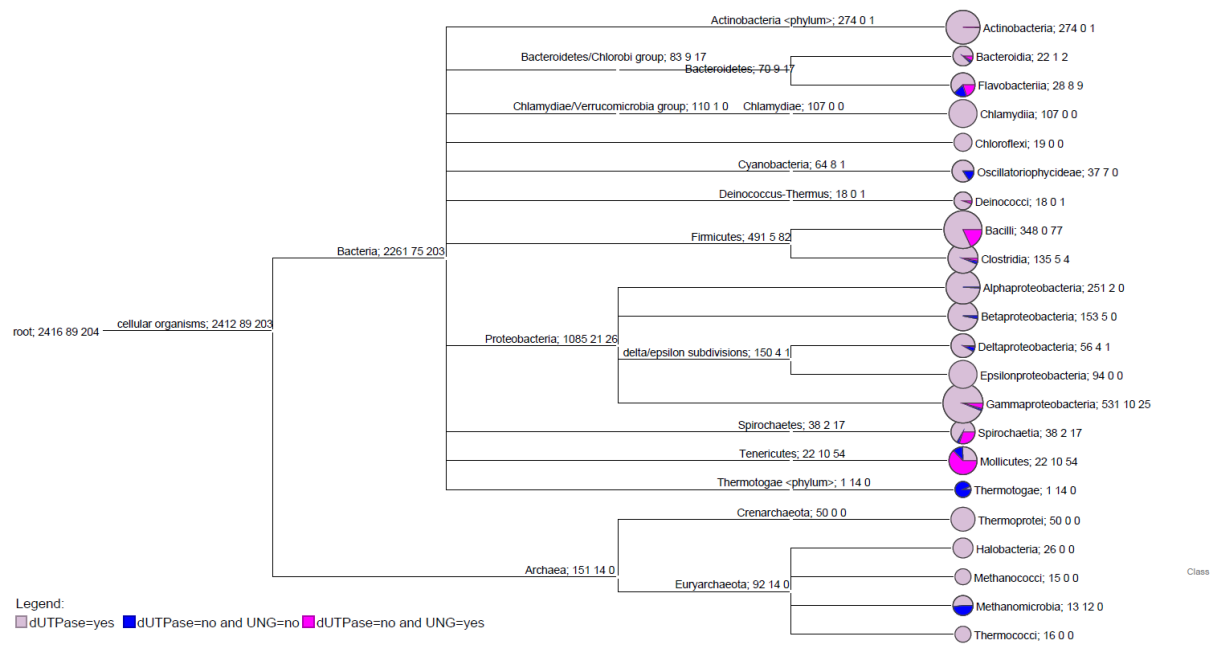

*Figure 2. The distribution of bacterial/Archaeal genomes without dUTPase. Only those classes are shown that have at least 15 genomes examined. Each node of the tree is labelled by three numbers: the first is the number of genomes with dUTPase under the node; the second is the number of genomes without dUTPase and UNG; the third is the number of genomes without dUTPase and with UNG. Since we show only the classes with at least 15 genomes at the right, the not shown classes account for the genomes, missing from the summation.*

**Materials and Methods**

Here we describe the workflow that has generated the list of bacterial and Archaeal genomes without dUTPase and from these genomes those with and without UNG, UGI, SAUGI and P56. The list, tables and the source of the in-house programs referred below, are available at the website http://pitgroup.org/static/life_wo_dutpase/.

*Finding bacterial genomes that do not contain dUTPase*

The source of the bacterial and Archaeal genome sequences was downloaded from the NCBI FTP site: ftp://ftp.ncbi.nlm.nih.gov/genomes/Bacteria/all.fna.tar.gz . For sequence search and alignment, the stand-alone UNIX blast program (25) was applied from the site http://www.ncbi.nlm.nih.gov/books/NBK52640/ on our local servers. Next, with the `makeblastdb` program, databases were generated for the genomic sequences for processing with blast. We filtered out the DNA sequences corresponding to plasmids by applying our in-house scripts `GenAllGenomesFileNames.sh` and `allgenomes_wo-plasmids.pl`.
Search for dUTPase sequences, the UNG sequence and the UNG inhibitor UGI-SAUGI-P56 sequences were directed by the `run-blast.pl` script that calls the program `tblastn`; the applied fasta files to search for in the database were: `dUTPase-tri-di1-di2-arch.fasta, UNG.fasta, UGI-SAUGI-P56.fasta.`, all downloadable from http://pitgroup.org/static/life_wo_dutpase/.
The dUTPase fasta file contains one trimeric (*E. coli* dUTPase, UniProt: `P06968`), two dimeric (*C. jejuni* and *S. aureus* phiEta phage dUTPases, UniProt: `O15826` and `Q9G011`, respectively), as well as and one Archaeal dUTPase-like sequence (the putative dCTP deaminase from *Pyrococcus furiosus*, Uniprot accession number `Q8X251`). The UNG fasta file contains the NCBI Reference Sequence `WP_001262716.1` of Enterobacteriaceae uracil-DNA glycosylase. The fasta file for the UNG inhibitor proteins consists of the sequences corresponding to the UniProt accession numbers `P14739, Q936H5` and `Q38503`.

The evaluation of the `tblastn` results were performed by the script `find-nohits.pl` that returned a table of the bacterial/Archaeal genomes without dUTPase genes where no alignments were found with smaller than 0.01 E-value for any of the three dUTPases we search for. The genomes without dUTPase hits were also partitioned into classes (i) according to the containment of UNG genes with better than 0.01 E-value, and (ii) containment of any UNG inhibitors with sequence-similarities from the fasta file `UGI-SAUGI-P56.fasta` of 0.01 E-value or less. The genomes without dUTPase and with UNG are listed in Supplementary Table S1. The memberships in the partitions of (i) and (ii) are denoted in the first two columns of Table S1. The genomes without both dUTPase and UNG are listed in Supplementary Table S2.

The interested reader can easily reproduce the results in each row of Tables S1 and S2 by using the on-line webserver at NCBI at the site:
http://blast.ncbi.nlm.nih.gov/Blast.cgi?PROGRAM=tblastn&PAGE_TYPE=BlastSearch&LINK_LOC=blasthome by choosing the "*Align two or more sequences*" option, copying the content of the fasta file `tri-di1-di2-arch-UNG-UGI-SAUGI-P56.fasta` in the first and copying the NC number of the row of the table into the second input field, and setting "Expect threshold" value to 0.01 at the "Algorithm parameters" menu (see the Supplementary Figure S2 for a screenshot). The hits are colored black while the sequences without hits by gray color.

*Generating the taxonomic distribution figure from the results Tables S1 and S2:*

We have used the MEGAN5 (26) metagenomic analysis software in a creative way for generating the evolutionary distribution of the genomes with and without dUTPase and UNG. Certainly, we do not have metagenomes here, but we can exploit a particular capability of the MEGAN5 software as follows. MEGAN5 is capable of comparing the taxonomic distribution of three metagenomes, and it can generate a phylogenetic tree to visualize the distribution. The membership in the three metagenomes can be described by a length-3 0-1 characteristic vector, the $i^{th}$ value is 0 if the taxon is not in the metagenome and 1 if it is in the metagenome, for $i$=1,2,3. Here we substitute these "memberships in metagenomes" with the memberships of sets of genomes with and without dUTPase and UNG as follows: 1,0,0 is substituted if the genome contains dUTPase gene, 0,1,0 is written if the genome does not contain dUTPase but it contain UNG, and 0,0,1 is written if the genome does not contain dUTPase and UNG.

The more technical description of the workflow is as follows.

First, the file that maps the gi values the Taxonomy IDs was downloaded from the NCBI FTP site: ftp://ftp.ncbi.nlm.nih.gov/pub/taxonomy/gi_taxid_nucl.dmp.gz. From this file, using the non-plasmid bacterial/Archaeal genome-headers, with a script enclosed as `Annot-w-TAXID.pl`, NC-numbers were mapped to `gi` and Taxonomy IDs; the resulting file is `NC-GI-TAXID-wo-plasmid.csv`.

Next, the gen-megan.pl script of ours was applied to get `life_wo_di1-di2-tri-arch_dUTPase_E001.megan` file that was opened by the MEGAN5 software (downloadable from http://ab.inf.uni-tuebingen.de/software/megan5/. The evolutionary tree figures were created by setting the Rank, and in the Tree menu by setting the `Show Number of Read Summarized` and `Show values on log scale` options. The leaves, containing only few genomes can be filtered by setting the `Tree/Hide Low Support Nodes` option in MEGAN5.

**Results and Discussion**

Figure 1 describes how UNG and dUTPase collaborate to keep DNA uracil-free and also shows the inhibitory protein factors described so far in the literature for either dUTPase or UNG. To date, only one dUTPase-inhibitory protein has been identified at the molecular level, namely the repressor protein termed Stl. This protein is encoded within the *S. aureus* SaPIBov1 pathogenicity island. For UNG, three different proteins have been identified with significant inhibitory effectivity. Two of these (UGI and p56) are encoded by different bacteriophages (phages PBS1/PBS2 and phi29 of *Bacillus subtilis* ((27,28), respectively). The UGI function encoded in phages is either required to allow synthesis of uracil-enriched DNA (in the case of phages PBS1/PBS2) or protects against the cleavage of phage genome at uracil positions thereby facilitating viral DNA replication (29). The third protein with UNG inhibitory activity was recently identified in *S. aureus* (SaUGI) and interestingly, this is the first such case where a UNG inhibitor is encoded in the cellular genome itself (30).

Both dUTPase and UNG are generally presumed to be ubiquitous in free-living organisms. It was therefore an unexpected finding that in *S. aureus*, the dUTPase gene is only found located on phages or prophages inserted into the cellular genome, while in strains cured of prophages and phages, the dUTPase gene is absent from the genome (16). Such conditions where the dUTPase enzymatic activity is down-regulated or missing are highly deleterious but may be well tolerated if the uracil-DNA glycosylase activity is diminished. In light of the recent studies on dUTPase and UNG inhibitory proteins, we set out to investigate the genotypes of prokaryotes and Archaea and in these organisms, we describe the distribution of genes that act for or against of uracil occurrence in DNA.

In our studies, we investigated those prokaryote and Archaea genomes that are fully sequenced and deposited in the NCBI Genome database. For dUTPases, two protein families have been described to date, the all-beta trimeric and the all-alpha dimeric dUTPases (11), hence we used representative sequences of these families in our search (dUTPases from *E. coli* and *C. jejuni*, respectively). Some Staphylococcal phages also encode a variety of dimeric dUTPase, hence one such sequence was also inserted in the search. In addition, some dCTP deaminases, especially from Archaea, were shown to belong to the trimeric dUTPase fold and acting as bifunctional dCTP deaminase/ dUTPase enzymes. One such sequence was therefore also included (namely dCTP deaminase from *P. furiosus*). For uracil-DNA glycosylase, the sequence of the UNG enzyme from *E. coli* was used in our search, as this subfamily of uracil-DNA glycosylases is associated with the major uracil excising efficiency.

The result of screening the bacterial and Archaeal genomes for the presence/absence of dUTPase and UNG genes is shown in Figure 2. Interestingly, this systematic approach revealed that the lack of dUTPase genes is far more frequent than usually thought. Numerous evolutionary branches showed up where a few or more species do not encode dUTPase protein (note the colored segments on Figure 2). In fact, most of the phyla contained some species where the dUTPase genes were not found. These instances are widely occurring on the bacterial evolutionary tree, and also among Euryarchaeota. These cases were further distributed into two groups depending on the simultaneous absence or presence of UNG gene (cf blue and pink segments on Figure 2, respectively). These two groups are expected to constitute highly different physiological conditions. Dual lack of both dUTPase and UNG possibly results in a viable phenotype with uracil enrichment in the DNA while lack of dUTPase and presence of UNG is expected to result in genomic instability, and in many cases, cell death.

A more detailed analysis of the evolutionary distribution of species that do not have dUTPase genes is shown on Figure S1 (cf also Table S1 and S2). Table I summarizes those evolutionary groups where the occurrence of *dut-* genotypes is detected in >5% of all genomes within the given evolutionary group and also indicates if the UNG gene is present or absent.

*Table I.* **Distribution of dut- genotypes among bacteria and Archaea.**
*Evolutionary branches where the dut-ung+ or dut-ung- genotype occurs in >5% of all genomes within the given evolutionary group*

| *dut – ung+* | *dut – ung –* |
|---|---|
| Staphylococcaceae | Oscillatoriophycideae |
| Flavobacteriaceae | Thermoanaerobacterales |
| Bacillaceae | Oceanospirillales |
| Enterococcaceae | Mycoplasmataceae |
| Vibrionaceae | Thermotogaceae |
| Spirochaetaceae | Methanomicrobia |
| Mycoplasmataceae | |

In summary, despite the usual textbook knowledge, we have clearly demonstrated that dUTPase is far from being ubiquitous in prokaryotes and Euryarcheota. It is of immediate further interest to understand how the different organisms may cope with this unexpected situation, especially when UNG is still present.

Inhibitory proteins of UNG may modify the physiological scenario, hence we investigated if any of the UNG inhibitory proteins may be encoded in those bacterial and Archaeal genomes that showed up as *dut-ung+* in our analysis. We found that none of the phage-related UGI or p56 protein genes could be located on the genomes investigated. The gene for SaUGI, the *S. aureus* UNG inhibitory protein was located on the *S. aureus* genome, and a similar sequence was also found on the Butyrivibrio proteoclasticus genome but not elsewhere. Hence, uracil-DNA metabolism basically remains to be governed by the dUTPase and UNG enzymes, with only a very few exceptions, mostly *S. aureus* strains.

**Survival strategies and possible physiological consequences**

Since the *dut-ung+* genotype is expected to result in genomic instability, it was of interest to investigate if any specific strategy may be employed by the species that are characterized with this unusual feature. First of all, it is important to mention that for S. aureus, numerous phages have been described that encode dUTPase (representatives from either the all-beta trimeric or the all-alpha dimeric dUTPase enzyme families). It has been also described that in *Salmonella enterica*, the *S. enterica* Serovar Typhimurium Myophage Maynard also encodes a bona fide dUTPase gene (31). Although fully genomic sequence information is limited for other Salmonella phages, this specific instance of phage-encoded dUTPase in the Myophage Maynard indicates the possibility that Salmonella strains also rely on phage-provided dUTPases.

Another strategy to supply some dUTPase-like enzymatic activity was found in *Deinococcus radiodurans*. This organism, known for its high resistance against ionizing radiation (32), encodes a MazG-like enzyme, with a rather promiscuous substrate specificity (33). Among numerous dNTPs, the MazG-like *D. radiodurans* enzyme also cleaves dUTP (33). Although less efficient and less specific, this supplementation of dUTPase enzymatic activity may ensure viability. In this respect, it is relevant to point out that in several systems, strong inhibition of dUTPase did not lead to lethality indicating that a residual dUTPase activity might be still enough for survival (12,34). Under these circumstances, the genomic DNA may contain a somewhat elevated level of incorporated deoxyuridine moieties.

For Thermatoga and Methanomicrobia, data from the literature indicate that the *dut-ung-* genotype found in our present work may be compensated for by including genes for a less specific MazG-like dNTPase together with an Archaea-like uracil-DNA glycosylase (35). Lateral gene transfer between Archaea and bacteria has been suggested as the underlying mechanism that led to the appearance of Archaea-like uracil-DNA glycosylase in Thermatoga.

In conclusion, we have shown that the genes for the common dUTPase enzyme families are far from being ubiquitous in prokaryotes and Archaea. This unexpected genotype is observed in evolutionary well-separated branches suggesting that loss of the *dut* gene(s) might have occurred on multiple independent occasions during evolution.

**Supplementary tables and figures**

The supplementary material is downloadable from
http://uratim.com/Life_without/LWO_Supplementary.zip

Figure S1 depicts the taxonomic distribution of bacterial/Archaeal genomes without dUTPase on the family level. Only those families are shown that have at least 15 genomes examined. Each node of the tree is labelled by three numbers: the first is the number of genomes with dUTPase under the node; the second is the number of genomes without dUTPase and UNG; the third is the number of genomes without dUTPase and with UNG. Since we show only the families with at least 15 genomes at the right, the not shown classes account for the genomes, missing from the summation. Blue color denotes the proportion of genomes without dUTPase and UNG, while pink genomes without dUTPase and with UNG.

Figure S2 is a screenshot showing the proper settings for the verification of our results with the NCBI tblastn webserver.

Table S1 gives the list of the bacterial/Archaeal genomes without dUTPase but with the UNG gene. The second column shows the presence of UNG inhibitors in the genome.

Table S2 gives the list of the bacterial/Archaeal genomes without dUTPase and UNG.